\documentclass[reprint,superscriptaddress,showpacs,amsmath,amssymb,aps,prl]{revtex4-2}

\usepackage{graphicx}
\usepackage[colorlinks=true,allcolors=blue]{hyperref}

\begin{document}

\title{Audio Classification with Skyrmion Reservoirs}

\author{Robin Msiska}
\affiliation{%
 Faculty of Physics, University of Duisburg-Essen, 47057 Duisburg, Germany
}%

\author{Jake Love}
\affiliation{%
 Faculty of Physics, University of Duisburg-Essen, 47057 Duisburg, Germany
}%

\author{Jeroen Mulkers}
\affiliation{
DyNaMat, Department of Solid State Sciences, Ghent University, 9000 Ghent, Belgium
}

\author{Jonathan Leliaert}
\affiliation{
DyNaMat, Department of Solid State Sciences, Ghent University, 9000 Ghent, Belgium
}

\author{Karin Everschor-Sitte}%
\affiliation{%
 Faculty of Physics, University of Duisburg-Essen, 47057 Duisburg, Germany
}%
\affiliation{%
Center for Nanointegration Duisburg-Essen (CENIDE), 47057 Duisburg, Germany
}%

\date{\today}

\begin{abstract}
Physical reservoir computing is a computational paradigm that enables spatio-temporal pattern recognition to be performed directly in matter.  The use of physical matter leads the way towards energy-efficient devices capable of solving machine learning problems without having to build a system of millions of interconnected neurons. We propose a high performance ``skyrmion mixture reservoir'' that implements the reservoir computing model with multi-dimensional inputs. We show that our implementation solves spoken digit classification tasks at the nanosecond timescale, with an overall model accuracy of 97.4\% and a $<$1\% word error rate -- the best performance ever reported for in-materio reservoir computers. Due to the quality of the results and the low power properties of magnetic texture reservoirs, we argue that skyrmion fabrics are a compelling candidate for reservoir computing.
\end{abstract}

\maketitle

\section{\label{sec:introduction}Introduction:\protect}
Reservoir computing (RC) is a computational paradigm inspired by the framework of recurrent neural networks that uses dynamical systems (reservoirs) to perform pattern recognition. Initial experiments in RC revolved around the usage of artificial neural network reservoirs, such as echo state networks~\cite{jaeger2001} and liquid state machines~\cite{maass2002}, where the reservoir consists of a large number of individual interconnected nodes or neurons. More recently, it has been shown that physical substrates can also provide the dynamics required for RC~\cite{tanaka2019}, provided they are sufficiently complex, non-linear and have a `fading memory' property~\cite{dambre2012}. Using a physical system over neural network-based models allows one to take advantage of the inherently non-linear nature of matter and realise energy-efficient computations without being constrained by the difficulties of implementing highly connected networks of neurons.

Physical reservoir computers consist of two main components, a non-linear substrate, known as `the reservoir', and a trainable linear readout layer. Temporal inputs are fed into the substrate in the form of physical perturbations, exciting the state of the reservoir. By observing a high-dimensional finite representation of the substrate, it is possible to construct a mapping from the original temporal input to a high-dimensional latent space in which input features can be classified linearly. For an ideal reservoir, one can use the results of previously encountered mappings to train the readout layer to recognise patterns in individual input signals via a linear model. Various examples of physical reservoir computing have been successfully implemented in a diverse set of substrates~\cite{dale2016, fernando2003, tanaka2019} including spintronics~\cite{torrejon2017, nakane2018} and skyrmion based systems~\cite{prychynenko2018, pinna2020, vedmedenko2020, dale2021, yokouchi2022, lee2022, raab2022}. Skyrmion and other spintronics-based reservoirs are of particular interest due to the ability to integrate into exis\-ting CMOS devices~\cite{grollier2020,finocchio2021}, their efficient response to low power excitations~\cite{lukovsevivcius2009},
and their tuneable properties for performance optimisation for a diverse set of problems~\cite{love2021}.

\begin{figure}
     \centering
     \includegraphics[width=\linewidth]{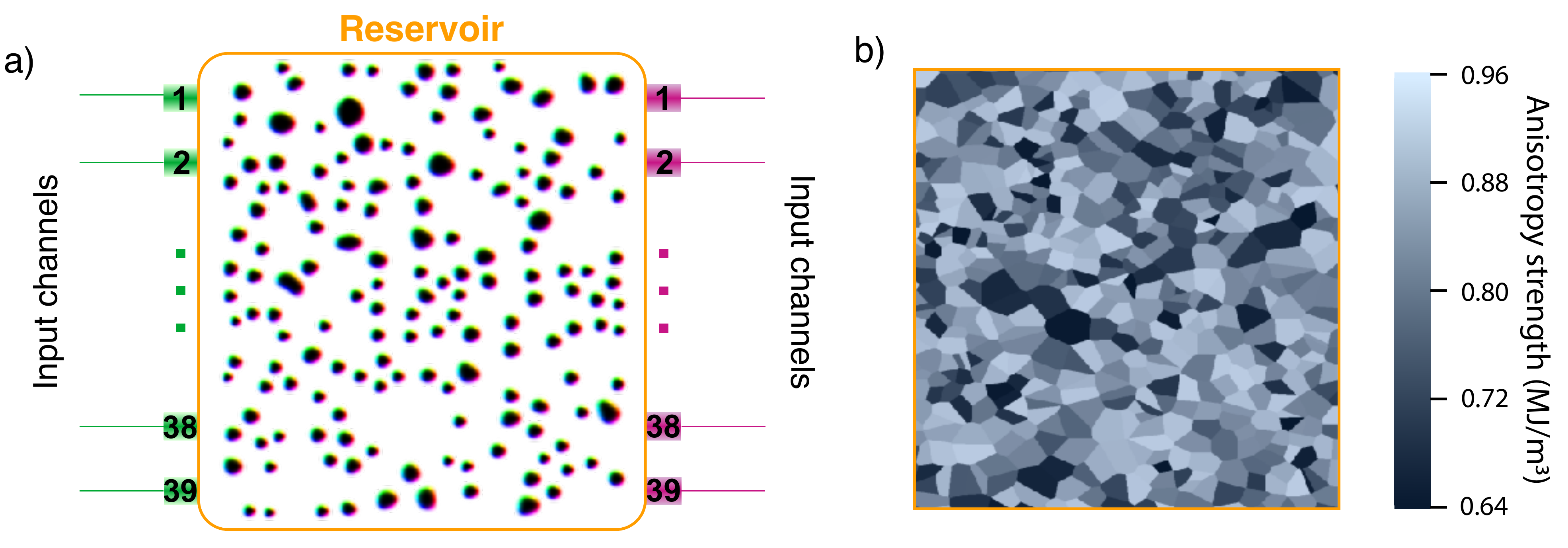}
     \caption{a) Schematic of multi-contact skyrmion reservoir with 39 electrical contacts on each side. The dark points represent skyrmion cores and the cyclic colourmap around them represents the in-plane magnetisation~\cite{vansteenkiste2014}. Contacts have been enlarged for visibility.  b) Random grain distribution underlying the skyrmion reservoir. Each grain has a particular anisotropy strength.}
     \label{fig:reservoir}
\end{figure}

The skyrmion-based reservoir computers considered so far operate with single input channels.
This paper presents a skyrmion RC model capable of nanosecond timescale pattern recognition for multi-dimensional inputs. 
In contrast to single input channel RCs, multi-channel input systems provide the advantage of having lower error rates and better power efficiency~\cite{katumba2017}. In this work, we firstly introduce the multi-channel
skyrmion reservoir model. The performance of our multi-channel skyrmion reservoir is benchmarked using a spoken digit speech recognition task. We show that we can solve the task to a high model accuracy of 97.4\% --- the highest ever reported for in-materio reservoir computers. We also performed the classification benchmark with a subset consisting only of female speakers, as typically done in RC research~\cite{appeltant2011, torrejon2017, abreu_araujo2020, welbourne2021}, and found a model accuracy of 98.5\%; a result which is on par with the highest accuracies reported in physical reservoirs. 

\begin{figure*}
     \centering
     \includegraphics[width=\linewidth]{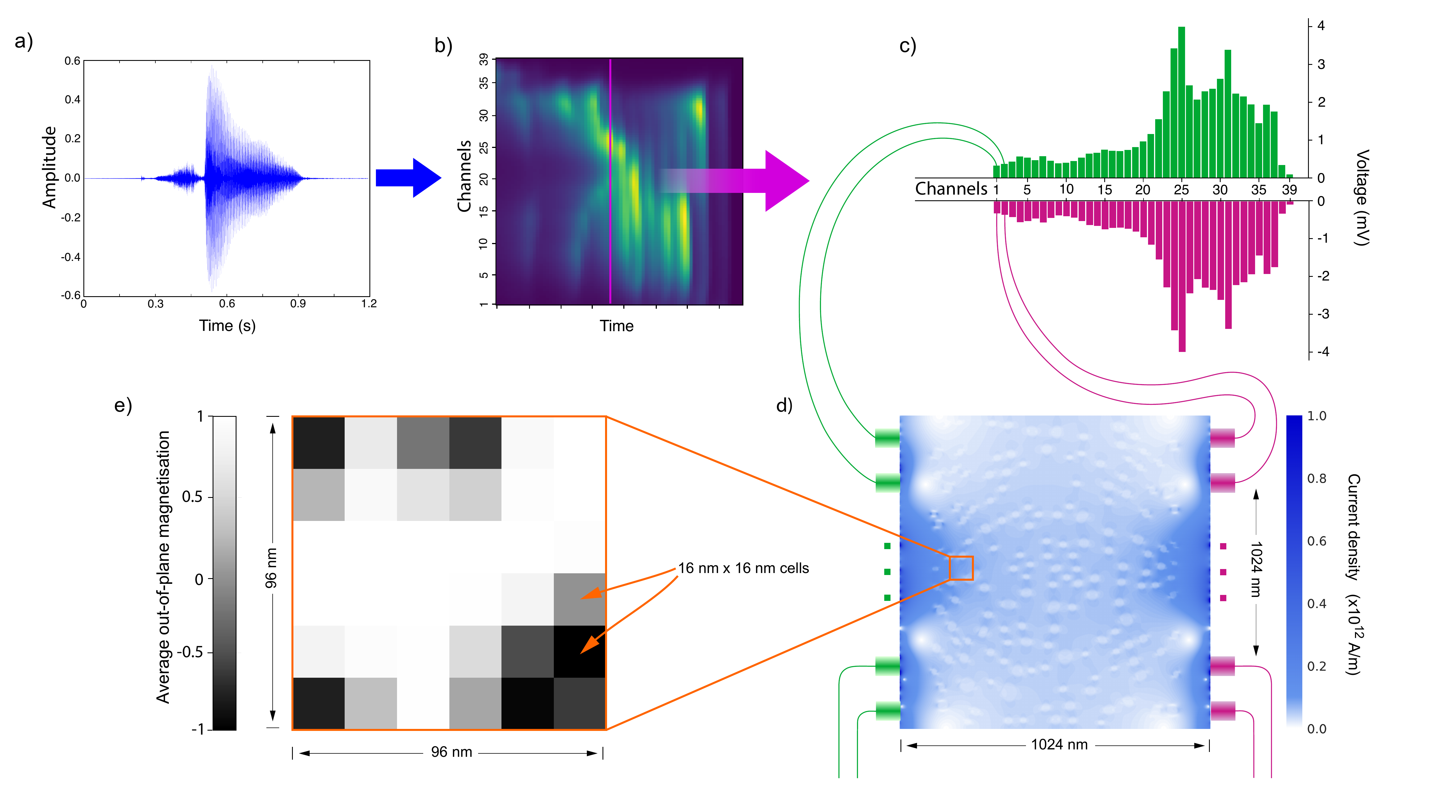}
     \caption{Skyrmion-based reservoir computing scheme used for audio recognition. a) An audio waveform is first created from a spoken word recording and then converted to b) a cochleagram. For every given time step (example indicated by the purple line), the frequency channels of the cochleagram are converted to c) voltage signals corresponding to the frequency intensities of each channel. These voltages are fed into d) a multi-contact skyrmion reservoir, shown here with a resolving current pattern depicted by the colourmap. The contacts on the reservoir are enlarged for visibility. A 96 nm $\times$ 96 nm section (orange box) of the skyrmion reservoir is enlarged in e) showing the 16 nm $\times$ 16 nm cells in which the out-of-plane magnetisation is averaged. Resolving magnetisation this way gives the output readout of the reservoir.}
     \label{fig:audio_process}
\end{figure*}

\section{Multi-contact skyrmion reservoir:\protect\label{sec:model}}

In this study, we consider a skyrmion reservoir made up of a magnetic thin-film hosting skyrmions in a random configuration. To this end, we attach 78 electrical contacts evenly spaced and aligned vertically over two opposite edges of the film such that each edge has 39 contacts. Figure~\ref{fig:reservoir}a) shows a schematic of this system. Magnetic thin films, such as the ones hosting the skyrmion reservoir, are typically polycrystalline materials made up of many grains with sightly varying material parameters~\cite{leliaert2014}. Our proposed skyrmion reservoir has a random assortment of grains, see Figure~\ref{fig:reservoir}b), each with a specific anisotropy strength. This leads to inhomogeneities in our system that favour certain regions of the film to accommodate skyrmions. 
Effective pinning of skyrmions~\cite{gruber2022} is induced by the grains and consequently, certain arrangements of skyrmions become meta-stable states to which the system can relax following transitory excitations  --- a feature that ensures the echo state property in such a reservoir. More details on the reservoir can be found in the methods section.

We exploit the magnetic reservoir's response to spatiotemporal electrical excitations to perform RC tasks such as spoken digit classification.

\section{Spoken Digit Classification:\protect\label{sec:results}}

To demonstrate the high-quality performance of the reservoir at solving multi-dimensional classification tasks, we subject it to a canonical benchmark task of audio recognition in which we classify isolated English spoken digits (from 0 to 9). We use the full set of spoken digits from the audio recordings from the TI-46 dataset~\cite{liberman1993}.  Figure~\ref{fig:audio_process}a) shows an example of an audio waveform of the raw data.
We pre-process the audio waveform signals using the Lyon ear model~\cite{lyon1982}, which converts each audio sample into a multi-dimensional signal in the form of a special type of spectrogram called a cochleagram, see Fig.~\ref{fig:audio_process}b). The cochleagram filters each audio time step into 39 frequency channels. At a given instance in time, each channel's frequency intensity is converted to voltage signals, see Fig.~\ref{fig:audio_process}c). These time-varying potentials
serve as inputs to the reservoir. On one side of the reservoir, we apply the corresponding input channel voltage to each contact. To each contact on the opposite side of the reservoir, we apply the negative value of the matching channel voltage. This creates an electrical potential with a steeper gradient than if contacts on one edge were grounded.

The skyrmion fabric making up the reservoir manifests an anisotropic magnetoresistive response to the voltage pulses at the contacts~\cite{bourianoff2018}. Resultant current densities interact with the magnetic skyrmions in the film through spin-torques and lead to distinctive time-varying skyrmion deformations.  Due to skyrmion pinning, an input-dependent current density arises~\cite{pinna2020}, see Fig.~\ref{fig:audio_process}d). In this manner, electrical excitations echo the inputs, thus allowing for reservoir computing. 

We employ a spatially resolved readout method on the output states of our reservoir system, with a resolution that is close to the size of the individual skyrmions considered in this work. This is done by partitioning the reservoir film into 64 $\times$ 64 non-overlapping discrete cells, as shown in Fig. \ref{fig:audio_process}e). For an overview of skyrmion diameters in different materials, see Refs.~\cite{everschor-sitte2018,tokura2021}. This readout technique captures regions of the reservoir that exhibit emergent magnetisation dynamics which characterise the reservoir’s computational properties. 

\begin{figure}
     \centering
     \includegraphics[width=1\linewidth]{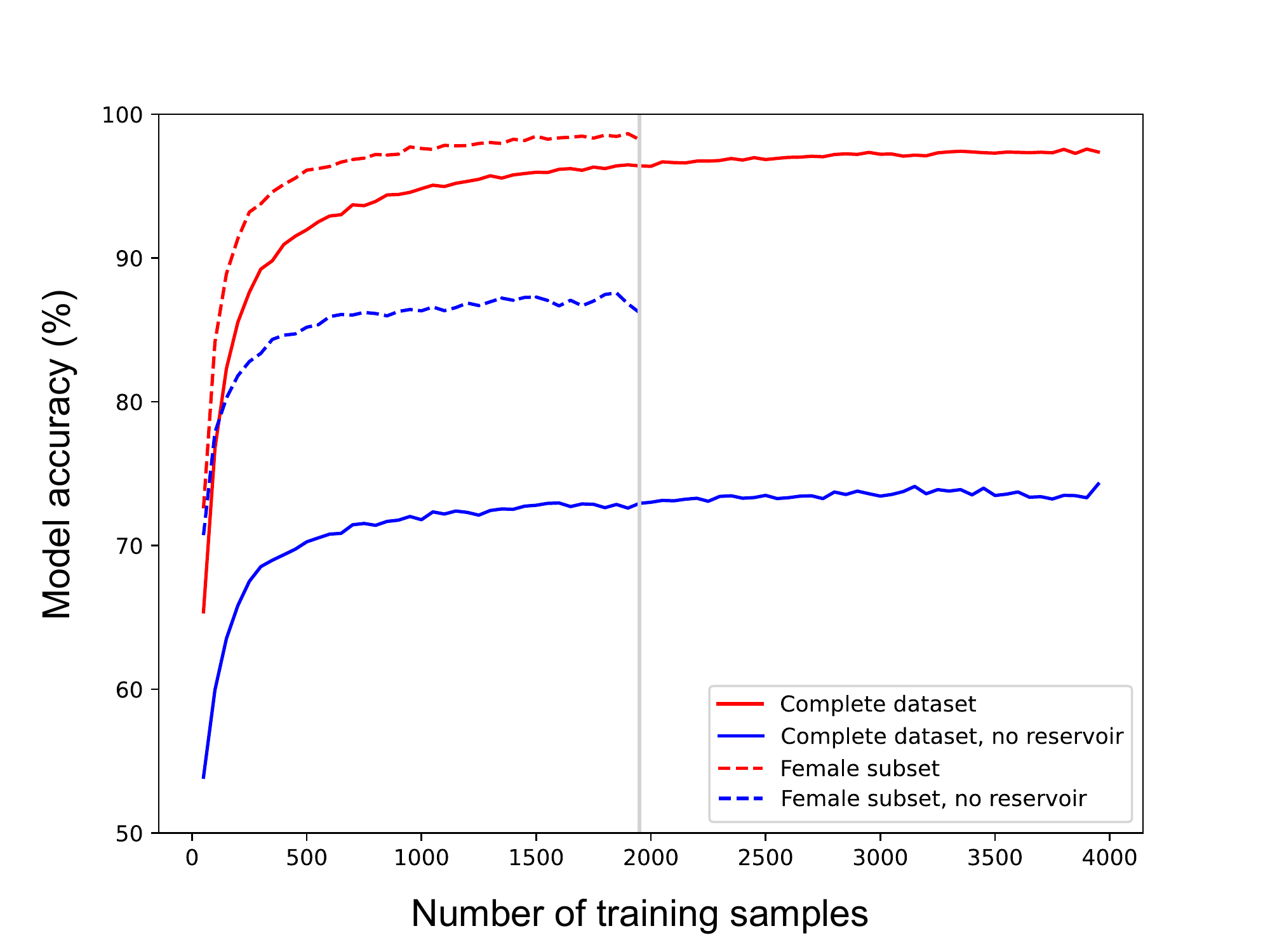}
     \caption{Accuracy curves from the training of the regression model, with (red) and without a reservoir (blue) for the complete (solid lines) and female only dataset (dashed lines).}
     \label{fig:accuracy}
\end{figure}

\subsubsection{Model training}
For the skyrmion reservoir, we create a linear model from the readout node (magnetisation) values. The RC is trained using supervised training via ridge regression. We use training curves and the word error rate (WER) metric to quantify our reservoir's performance. Training curves evaluate changes in error with increasing training set size while WER evaluates the percentage difference between predicted output labels  and true output labels.

 The dataset used contains spoken digits from 16 speakers (8 male and 8 female), each providing 26 utterances of numbers 0 through 9. This gives us 4160 total samples which we split into about 85\% training \& validation data and 15\% test data. We test the efficacy of our RC scheme by comparing classification results in the presence and in the absence of a reservoir. We standardise the training/testing data of our regression model to ensure that regularization is not biased by the possible difference in units when the reservoir is absent~\footnote{Note that when we refer to the absence of a reservoir we mean that we create a regression model directly from the data in the channels of the cochleagram, such as the one depicted in Fig. \ref{fig:audio_process}b).} vs. when it is present. Furthermore, we also create a linear model from a subset of the audio dataset consisting of only utterances from female speakers, in order to compare our model accuracy to existing works~\cite{appeltant2011, torrejon2017, abreu_araujo2020, welbourne2021}.

Fig.~\ref{fig:accuracy} shows training accuracy curves for the complete dataset as well as for female speakers only, with and without a reservoir. 
 The skyrmion reservoir yields the highest accuracy of around 98.5\% for the female subset. The trained model for the complete data produces 97.4\% accuracy. 
 For both cases, the accuracy is much lower without the reservoir. 
 With our models trained, we can perform predictions on the test datasets.
 
\begin{figure}
     \centering
     \includegraphics[width=0.5\textwidth]{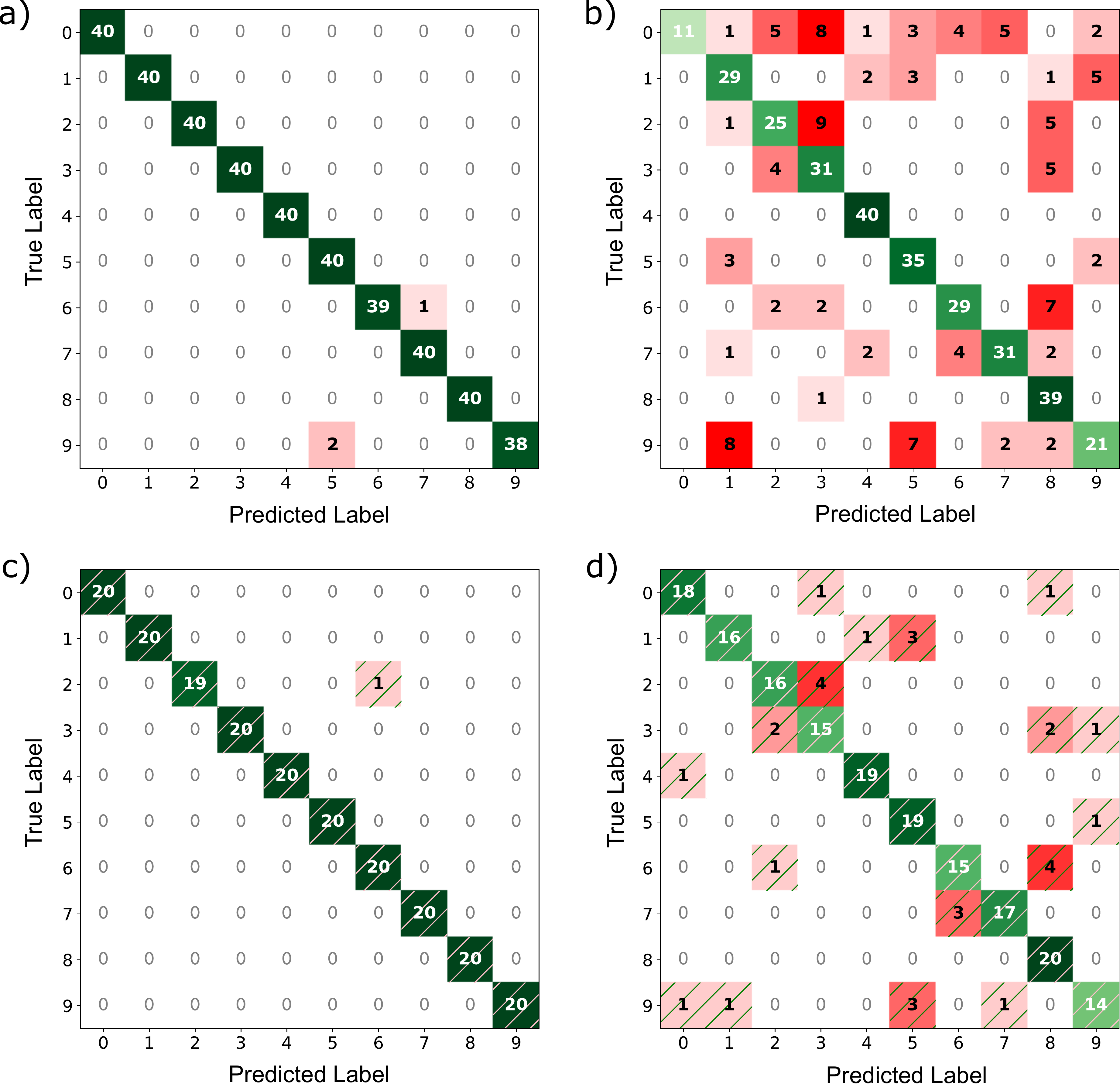}
     \caption{Classification performance visualised by confusion matrices. Correctly predicted utterances lie on the diagonal. The colour intensity reflects the relative number of correctly (green) and incorrectly (red) predicted utterances. The solid coloured plots show results for the full spoken digit dataset a) with and b) without skyrmion reservoir. The hatched plots denote the confusion matrices for the female subset c) with and d) without skyrmion reservoir.}
     \label{fig:confusion}
\end{figure}

\subsubsection{Model testing}
 Classification performance is quantified using a confusion matrix, with the predicted labels as the columns and the true labels as the rows. The resulting confusion matrices for the full dataset and the female subset are shown in Fig.~\ref{fig:confusion}. 
 We define WER as the ratio of incorrectly predicted versus the total number of utterances. Applying this notion of WER to the confusion matrices in Fig.~\ref{fig:confusion}, we obtain WER values shown Tab.~\ref{tab:WER}. 
 
\begin{table}[h]
\begin{tabular}{|c|c|c|} 
\cline{2-3}
\multicolumn{1}{l|}{} & With reservoir     & Without reservoir  \\ 
\hline
\rule{0pt}{4ex} Full dataset                & 3/400 $\rightarrow$ 0.75\%   & 109/400 $\rightarrow$ 27.25\%    \\ 
\hline
\rule{0pt}{4ex} Female subdataset               & 1/200 $\rightarrow$ 0.50\%  & 31/200 $\rightarrow$ 15.50\%   \\
\hline
\end{tabular}
\caption{Word error rates for the full and female-only dataset, with and without the skyrmion reservoir.
\label{tab:WER}
}
\end{table}
Similar to the results of model training, we find that the skyrmion reservoir enhances greatly the classification performance and reduces the WER to 0.75\% (0.5\%) for the full (female only) dataset. Note that the reason we only have 200 testing samples for the  female subdataset is because it contains half as many samples as the full dataset.

\section{Discussion
\protect\label{sec:discussion}}

The quality of the results presented highlights the benefits of employing a multi-channel scheme for RC.  Such an architecture allows for greater mixing of signals thus adding more memory to the reservoir as well as nuance to output states \cite{katumba2017}. Furthermore, multiple input contacts ensure that currents access more regions of the sample, thereby reducing the total noise of the system.

In our skyrmion reservoir simulations we considered systems with the material parameters specified in the methods section, at zero temperature. Although different material parameters will affect the quantitative results, the specific choice of the material parameters, which play the role of hyper-parameters in the context of RC, are inconsequential for the qualitative results presented in this work. The extent to which thermal effects influence the system depends on many aspects, such as sample thickness, material properties etc.~\cite{pinna2020, litzius2020}. In specially designed, low-pinning materials, skyrmions have been shown to perform thermally driven diffusive motion on the millisecond timescale~\cite{zazvorka2019}. As the operation of our skyrmion reservoir is based on the motion of the skyrmions within the potential wells of the material grains on the GHz timescales of the input, we do not expect thermal fluctuations to significantly negatively affect the performance.

In this work we presented the results for RC spoken digit benchmark classification using the full TI-46 spoken digit dataset as well as the female subset.
For the latter we used all eight female speakers for which our RC performed extraordinarily well with a model accuracy of 98.5\% and a 0.5\% error rate. 
A common practice in the literature, however, is to create a subdataset of TI-46 consisting only of recordings of the first five female speakers. For a one-to-one comparison of our reservoir with such works~\cite{appeltant2011, torrejon2017, abreu_araujo2020, welbourne2021}, we benchmarked on this sub-subset and we obtained a 98.6\% model accuracy and a 0\% WER for prediction. Hence, our female subset results are of the same calibre as state-of-the-art reservoir systems including purpose-made electronic and optoelectronic reservoirs~\cite{vandoorne2014,zhong2021,abreu_araujo2020}. 
For the sake of completeness, we performed the classification on a male-only subdataset, and this yielded similar results as shown in Fig.~\ref{fig:confusion}c) and Fig.~\ref{fig:confusion}d). We obtained a model accuracy of 98.4\% and WER of 0.5\%.

\section{Summary:
\protect\label{sec:summary}}
We reported the first demonstration of spoken digit classification using a multi-channel skyrmion fabric reservoir computer. Using a non-linear filtering technique, we created a model able to identify each digit independent of the speaker with an overall accuracy of 97.4\% (98,5\% for a female subset), and an associated WER of less than 1\% for prediction. This performance is the highest ever reported for physical reservoir systems using the TI-46 benchmarking dataset. 

The high quality of the results shown, combined with the ease of obtaining multi-channel inputs for skyrmion-based reservoirs, provides a path for such RCs to efficiently solve complex spatiotemporal problems with multidimensional information content.

\begin{acknowledgments}
We acknowledge funding from the Emergent AI Centre funded by the Carl-Zeiss-Stiftung and the Deutsche Forschungsgemeinschaft under Project No. 320163632. J.L. was supported by the Fonds Wetenschappelijk Onderzoek (FWO-Vlaanderen) with senior postdoctoral research fellowship Nr. 12W7622N. Part of the computational resources and services used in this work were provided by the VSC (Flemish Supercomputer Center), funded by Ghent University, the Research Foundation Flanders (FWO) and the Flemish Government – department EWI. 
\end{acknowledgments}

\section{Methods:\protect\label{sec:methods}}
To simulate the dynamics of the skyrmion reservoirs, we use MuMax3~\cite{vansteenkiste2014} with a custom extension to calculate self-consistent currents in the presence of anisotropic magnetoresistive effects. The reservoir is modelled as a 1024 nm $\times$ 1024 nm magnetic film using a $1024\times1024\times1$ grid of cells such that the cell size is smaller than the exchange length of the material. The material parameters are chosen to mimic the Co layer in a Pt/Co/MgO system~\cite{wang2018}. The parameters are as follows: exchange stiffness $A_\text{ex} = 15$ pJ/m, DMI strength $D = 3.3$~mJ/m$^2$  and saturation magnetisation $M_s=580\ \mathrm{kA/m}$. Inhomogeneities were introduced to the system the form of Voronoi grains such that each grain has a uniformly distributed anisotropy strength of K$_u$ = 0.80 $\pm$ 0.16 MJ/m$^3$.

Electrical contacts are added to the left and right edges of the simulation box. We implemented each of the 78 electrical contacts as a discrete simulation region with a square shape. We systematically scale these contact regions in such a way that they are equally sized and have a uniform spacing of 10~nm. 
Additionally, to avoid skyrmion dynamics driven by the very high localised current densities around the contacts, we implement a buffer region with a width of 100 nm measured from each of the edges (left and right) and extending towards the center of the film in which the magnetization remains uniform.

The magnetisation state of the reservoir was initialised by creating a skyrmion lattice with a uniform distribution of skyrmion separated by 25 nm. The texture was then relaxed until a metastable stable state was reached. To further relax the magnetic state, a normal discrete random signal with a variance of 8 mV was generated and fed into the reservoir at a sample rate of 5 GHz for a duration of 200 ns. The random signal was fed through all input contacts simultaneously, with the goal of removing initial instabilities in the system thus helping to enforce the fading memory property of the skyrmion reservoir. 

We used the Lyon passive ear model \cite{lyon1982} to decompose the spoken digit audio signals to cochleagrams. This decomposition allows us to extract the most useful acoustic features which are embedded in the frequency domain rather than in the amplitude of an audio waveform \cite{ doddington1981}. Cochleagram frequency intensities are mapped to voltages as discussed in the main text.

The linear model used to train the output weights was created using the Scikit-learn software library for the Python programming language.

\end{document}